\documentclass[submission]{eptcs}

\usepackage{breakurl}             
\input{pqmodel.sty}

\title{A Categorical Model for a Quantum Circuit\\ Description Language
(Extended Abstract)}
\author{Francisco Rios and Peter Selinger
 \institute{Dalhousie University\\Halifax, Canada}}

\def\titlerunning{A Categorical Model for a Quantum Circuit
  Description Language} 
\def\authorrunning{Francisco Rios and Peter Selinger}

\begin{document}
\maketitle

\begin{abstract}
  Quipper is a practical programming language for describing families
  of quantum circuits. In this paper, we formalize a small, but useful
  fragment of Quipper called {\em Proto-Quipper-M}. Unlike its parent
  Quipper, this language is type-safe and has a formal denotational
  and operational semantics. Proto-Quipper-M is also more general than
  Quipper, in that it can describe families of morphisms in any
  symmetric monoidal category, of which quantum circuits are but one
  example. We design Proto-Quipper-M from the ground up, by first
  giving a general categorical model of {\em parameters} and {\em
    state}. The distinction between parameters and state is also known
  from hardware description languages. A parameter is a value that is
  known at circuit generation time, whereas a state is a value that is
  known at circuit execution time. After finding some interesting
  categorical structures in the model, we then define the programming
  language to fit the model.  We cement the connection between the
  language and the model by proving type safety, soundness, and
  adequacy properties.
\end{abstract}

\section{Introduction}

Quipper is a functional programming language for quantum computing
{\cite{GLRSV2013-pldi,GLRSV2013-quipper}}.  What distinguishes Quipper
from earlier quantum programming languages, such as the quantum lambda
calculus {\cite{SV2009-qlambdabook}}, is that it is a {\em circuit
  description language}. This means two things: on the one hand, the
language can be used to construct quantum circuits in a structured
way, essentially by applying one gate at a time. On the other hand,
the completed circuits themselves become data, which can be stored in
variables and passed as components to subroutines, and on which
meta-operations (such as circuit transformations, gate counts,
inversion, error correction, etc) can be performed. These two levels
of description (gate level operations and meta-operations on entire
circuits) correspond quite closely to how many quantum algorithms are
specified in the literature, and therefore provide a useful high-level
paradigm for quantum programming.

Quipper is {\em practical}; it has been used to implement several
large-scale quantum algorithms, generating circuits containing
trillions of gates. However, there are some drawbacks.  For efficiency
reason, the original Quipper language was implemented as an embedded
domain-specific language within the host language Haskell. As a result
of mismatches between Quipper's type system and that of Haskell,
Quipper is not type-safe (there are some well-typed programs that lead
to run-time errors). In particular, Haskell is unable to enforce {\em
  linearity}, i.e., the requirement that a quantum state cannot be
used more than once (also known as the no-cloning property of quantum
information). Moreover, as an embedded language, Quipper has no formal
semantics; its behavior is only defined by its implementation. Giving
a formal semantics of Quipper would require giving a formal semantics
of Haskell, which is not feasible.

In this paper, we formalize a small, but useful fragment of Quipper
called {\em Proto-Quipper-M}. This fragment is a stand-alone
programming language (i.e., not embedded in a host language), and it
has its own custom type system and semantics.  This research fits into
a larger program of formalizing portions of Quipper, and we use the
name {\em Proto-Quipper} more generally to refer to any such
formalized languages. For example, another version of Proto-Quipper
has appeared in Ross's Ph.D. thesis {\cite{RossThesis}}.

An important concept that arises in Quipper, and also in related
settings such as hardware description languages, is the distinction
between parameters and state. To understand the distinction, it is
useful to keep in mind that Quipper is a language not just for
describing individual quantum circuits, but parameterized {\em
  families} of quantum circuits: for example, in the case of Shor's
algorithm, one circuit for each integer to be factored. A {\em
  parameter} is a value that is known at circuit generation time, thus
potentially picking out a member of the family of circuits. A {\em
  state} is a value that is known at circuit execution time, such as
the state of a qubit. Naturally, a state can depend on a parameter,
but since states are not known at circuit generation time, parameters
cannot be a function of states. 

Conceptually, Quipper's gate level operations operate on states (for
example, by applying a gate to a qubit), whereas its circuit level
operations operate on parameters. One fundamental feature of Quipper
is that there is only one kind of variable, which can hold both
parameters and states (or any combination thereof, such as a list of
qubits: here the length of the list is a parameter, and the individual
qubits are states). We rely on the {\em type system} to ensure that
parameters and states are used correctly. This distinguishes Quipper
from related approaches, such as the QWire language
{\cite{Paykin2017a}}, which enforce a strict separation between
parameters and states by essentially defining two separate programming
languages, one for parameters and one for states. The Quipper approach
is more flexible, because the ability to mix parameters and states
more freely gives the programmer access to useful abstractions, such
as quantum data structures (like lists of qubits), or even pairs of
entangled functions.

Our approach to the design of Proto-Quipper-M is from the ground up.
We start by defining a general categorical model of parameters and
state, and identifying a number of interesting categorical structures
in the model. We then define the programming language to fit the
model. One advantage of this approach is that our programming language
is almost ``correct by design''. In fact, as a result of our abstract
approach, Proto-Quipper-M is slightly more general than Quipper, in
the sense that it can describe families of morphisms of an arbitrary
monoidal category, rather than just quantum circuits. We give
computational meaning to the language by defining an operational
semantics. Finally, we establish the correctness of the operational
semantics by proving type safety, soundness, and adequacy properties.

\section{A cartesian model of parameters and state}\label{sec-simple}

In this section, we describe a simplified categorical model of
parameters and state, which will be convenient for introducing some
useful terminology, and will be a stepping stone toward the more
general model of Section~\ref{sec-mmxx}. The model presented here is
cartesian, rather than monoidal, and therefore could be used to model
a language for describing classical, rather than quantum circuits.
Nevertheless, several important notions will already be visible in
this model.

\subsection[The model {[2op,Set]}]{The model $\Setar$}\label{ssec-cartesian}
\label{ssec-setar-object-properties}

\begin{definition}
  Let $\Two$ be the category with two objects, called $0$ and $1$, and
  a single non-identity arrow $0\rightarrow 1$.  Consider the functor
  category $\Setar$. Explicitly, an object of this category is a
  triple $A=(A_0,A_1,a)$, where $A_0,A_1$ are sets and $a:A_1\to A_0$
  is a function. A morphism $f:A\to B$ is a commutative diagram
  \begin{equation}\label{eqn-morphism}
    \vcenter{\xymatrix@R-1ex@C-1ex{A_1 \ar[d]_{a}\ar[r]^{f_1} & B_1 \ar[d]^{b} \\
      A_0\ar[r]^{f_0} & B_0.}}
  \end{equation}
\end{definition}

We can think of an object $A=(A_0,A_1,a)$ as describing a family of
sets, as follows. For each $x\in A_0$, we define
\begin{equation}\label{eqn-ax}
  A_x = \s{ s\in A_1 \mid a(s) = x }.
\end{equation}
The set $A_x$ is called the {\em fiber} of $A$
over $x$. Up to isomorphism, the object $A$ is uniquely determined by
the family $(A_x)_{x\in A_0}$. We call the elements of $A_0$ {\em
  parameters}. We call the elements of $A_x$ {\em states}, and we say
that the state is {\em over} $x$. The elements of $A_1$ can therefore
be identified with pairs $(x,s)$, where $x\in A_0$ and $s\in A_x$. We
also call the pairs $(x,s)$ the {\em generalized elements} of $A$.

The requirement that the diagram {\eqref{eqn-morphism}} commutes is
exactly equivalent to the statement ``states may depend on parameters,
but parameters may not depend on states''. To see this, consider the
effect of a morphism $f:A\to B$ on a parameter-state pair $(x,s)$ as
defined above. Let $y=f_0(x)$ and $t=f_1(s)$. Then $y\in B_0$ and
$t\in B_1$. Moreover, by the commutativity of {\eqref{eqn-morphism}},
we have $y = f_0(x) = f_0(a(s)) = b(f_1(s)) = b(t)$. In other words,
for each $x\in A_0$, $f_1:A_1\to B_1$ restricts to a function
$f_x : A_x \to B_{f_0(x)}$. We write $f(x,s) = (y,t)$.  We note that
$y$ is only a function of $x$, because $y = f_0(x)$.  Therefore,
parameters may not depend on states.  On the other hand, $t$ is a
function of both $x$ and $s$, because $t=f_x(s)$. Therefore, states
may depend on parameters and states.

\begin{example}\label{exa-bool-bit}
  We define some particular objects of $\Setar$. Let
  $\Bool = (2,2,\id)$, where $2=\s{0,1}$ is a 2-element set and $\id$
  is the identity function. Let $\Bit = (1,2,u)$, where $1=\s{*}$ is a
  1-element set and $u:2\to 1$ is the unique function. In diagrams:
  \[ \Bool ~=~ \vcenter{\xymatrix@R-1ex@C-1ex{2\ar[d]^<>(.5){\id}\\2}} \qquad
  \Bit ~=~ \vcenter{\xymatrix@R-1ex@C-1ex{2\ar[d]^<>(.5){u}\\1}}
  \]
  The two generalized elements of $\Bool$ are $(0,0)$ and $(1,1)$,
  which we identify with ``false'' and ``true'', respectively.  The
  two generalized elements of $\Bit$ are $(*,0)$ and $(*,1)$, which we
  again identify with ``false'' and ``true''.

  So what is the difference between $\Bool$ and $\Bit$?  Informally, a
  boolean is only a parameter and has no state, whereas a bit is only
  state and has no parameters.  Note that there is an ``identity''
  function $f:\Bool\to\Bit$, mapping false to false and true to
  true. This function is given by the commutative diagram
  \begin{equation}\label{eqn-init}
    \vcenter{\xymatrix@R-1ex@C-1ex{2\ar[r]^{\id}\ar[d]_<>(.5){\id} & 2 \ar[d]^{u} \\ 2\ar[r]^<>(.5){u} & 1,}}
  \end{equation}
  and it satisfies $f(0,0) = (*,0)$ and $f(1,1) = (*,1)$. On the other
  hand, there exists no morphism $g:\Bit\to\Bool$ mapping false to
  false and true to true: the diagram
  \begin{equation}
    \vcenter{\xymatrix@R-1ex@C-1ex{2\ar[r]^{\id}\ar[d]_<>(.5){u} & 2 \ar[d]^<>(.5){\id} \\ 1\ar[r]^{?}
      & 2}}
  \end{equation}
  cannot be made to commute. Therefore, a boolean can be used to
  initialize a bit, but not the other way round. This precisely
  captures our basic intuition about parameters and state.
\end{example}

Generalizing the example of $\Bool$ and $\Bit$, we say that an object
$A$ is a {\em parameter object} if it is of the form $(A,A,\id)$, and
it is called a {\em state object} or {\em simple} if $A_0$ is a
singleton.  Note that $\Bool$ is a parameter object and $\Bit$ is a
state object. Informally, a simple object corresponds to a single
piece of data, rather than a parameterized family of data. Every
object $A$ is isomorphic to a sum of simple objects, namely,
$A\iso \sum_{x\in A_0} A_x$.

\subsection{A lambda calculus for parameters and state}

Since the category $\Setar$ is cartesian closed, we can interpret the
simply-typed lambda calculus in it. We can also add sum types, base
types such as $\Bool$ and $\Bit$, and basic operations such as
$\init:\Bool\to\Bit$, all of which have obvious interpretations in the
model. In this way, we obtain a very simple and semantically sound
lambda calculus for the description of boolean (non-reversible)
circuits. Moreover, the category $\Setar$ is co-complete, which allows
us to interpret inductive datatypes such as $\List(A)$ using initial
algebra semantics.

\section{A categorical model of circuit families}\label{sec-mmxx}

While the model of Section~\ref{ssec-cartesian} is useful for formalizing
some basic intuitions about parameters and state, it is not yet a good
model for describing families of quantum circuits.  The main issue is
that the model is cartesian, rather than monoidal. For example, there
exist morphisms $\Delta : A\to A\times A$ for all objects $A$,
including state objects. This is not appropriate if we want to
describe quantum circuits, where the no-cloning property prevents us
from duplicating quantum states.

To see how to generalize the model $\Setar$ to a monoidal setting,
recall from {\eqref{eqn-ax}} that the objects of $\Setar$ can be
equivalently described as pairs $(A_0, (A_x)_{x\in A_0})$, where $A_0$
is a set and $(A_x)_{x\in A_0}$ is a family of sets. Then a morphism
$f:A\to B$ can be equivalently described as a pair
$(f_0,(f_x)_{x\in A_0})$, where $f_0:A_0\to B_0$, and for each
$x\in A_0$, $f_x:A_x\to B_{f_0(x)}$.  We generalize this by
considering $A_x$ to be an object of a monoidal category instead of a
set.

\subsection{The category $\Mm$: generalized circuits}
\label{ssec-category-Mm}

Before designing a circuit description language, we should be more
precise about what we mean by a ``circuit''.  Rather than specifying a
particular class of circuits, for example as graphical representations
of sequences of gates, we take a more general and abstract point of
view: for us, a circuit is simply a morphism in a (fixed but
arbitrary) symmetric monoidal category.  We therefore assume that a
symmetric monoidal category $\Mm$ is given once and for all, and we
call its morphisms {\em generalized circuits}. From this point of
view, Proto-Quipper is simply a language for describing families of
morphisms of $\Mm$.

\begin{remark}\label{rem-meta}
  We will regard the morphisms of $\Mm$ as {\em concrete} data. We
  should imagine that the category $\Mm$ is equipped with additional
  ``meta-operations'', such as methods to print morphisms, determine
  their size or cost, invert them, and so on. These additional
  operations are external to the category $\Mm$, and take the form of
  set-theoretic functions such as $\size:\Mm(T,U)\to \N$,
  $\print:\Mm(T,U)\to\Document$, $\invert:\Mm(T,T)\to\Mm(T,T)$. Here
  $\Mm(T,U)$ denotes a hom-set of the category $\Mm$, and
  ``$\Document$'' denotes a set of printable documents. If such
  meta-operations are present, we regard them as fixed and given.
\end{remark}

\subsection[The category M-bar: state]{The category $\Mmx$: state}

As the first step in constructing our model, we choose a full
embedding $\Mm$ in some symmetric monoidal closed, product-complete
category $\Mmx$.  This can be done in some fixed, but arbitrary
way. For example, the Yoneda embedding has the required properties,
using the Day tensor for the monoidal structure
{\cite{Day70,MacLane91}}. However, we do not specify any particular
way of constructing $\Mmx$.

\begin{remark}\label{rem-abstract}
  Unlike the category $\Mm$, we regard the category $\Mmx$ as {\em
    abstract}. It is monoidal closed, so we will be able to form
  higher-order objects such as $(A\lolli B\x C)\lolli D$.
  However, we do not imagine morphisms between such higher-order
  objects as being concrete things that can be printed, measured, etc.
  Instead, we will only interact with such higher-order morphisms via
  the monoidal closed structure. In other words, the higher-order
  structure only exists as a kind of ``scaffolding'' to support
  lower-order concrete operations. By not specifying a particular way
  of constructing $\Mmx$, but only specifying its properties, we
  ensure that we will treat this category in the abstract, i.e., none
  of the theorems we will prove depend on any properties of $\Mmx$
  other than those that were stated.
\end{remark}

\subsection[The category M-double-bar: parameter]{The category $\Mmxx$: parameters}

We now define the category $\Mmxx$, which will serve as a model for
parameters and state, and which will be the main carrier of the
categorical semantics of our circuit description language.

\begin{definition}\label{def-mmxx}
  The category $\Mmxx$ has the following objects and morphisms:
  \begin{itemize}
  \item An object is a pair $A=(X,(A_x)_{x\in X})$, where $X$ is a set
    and $(A_x)_{x\in X}$ is an $X$-indexed family of objects of
    $\Mmx$. As before, we call $A_x$ the {\em fiber} of $A$ over $x$.
    We sometimes write $X=A_0$.
  \item A morphism $f:(X,(A_x)_{x\in X})\to(Y,(B_y)_{y\in Y})$ is a
    pair $(f_0, (f_x)_{x\in X})$, where $f_0:X\to Y$ is a function and
    each $f_x : A_x\to B_{f_0(x)}$ is a morphism of $\Mmx$.
  \end{itemize}
\end{definition}

This category is surprisingly rich in structure. We begin by stating
some of its most fundamental properties. It is perhaps not very
surprising that $\Mmxx$ has coproducts and a symmetric monoidal
structure. What is perhaps more surprising is that it is also monoidal
closed.

\begin{proposition}
  The category $\Mmxx$ has an initial object, given by
  $0=(\emptyset,\emptyset)$, where $\emptyset$ denotes both the empty
  set and the empty family. It also has coproducts, given by
  $A+B = (A_0+B_0, (C_i)_i)$, where
  $A_0+B_0=\s{(0,x)\mid x\in A_0}\cup\s{(1,y)\mid y\in B_0}$ is the
  disjoint union of sets, and $C_{(0,x)} = A_x$ and $C_{(1,y)} =
  B_y$.
  The category $\Mmxx$ also has infinite coproducts, defined in an
  analogous manner. Indeed, it is well-known that $\Mmxx$ is the free
  coproduct completion of $\Mmx$.
\end{proposition}

\begin{proposition}
  The category $\Mmxx$ is symmetric monoidal closed with the following
  structure:
  \[
  \begin{array}{lll}
    I &=& (1,(I))\\
    A\x B &=& (A_0\times B_0, (A_x\x B_y)_{(x,y)\in A_0\times B_0})
    \\
    A\lolli B &=& (A_0\to B_0, (C_f)_{f}),
  \end{array}
  \]
  where
  $
  C_f = \prod_{x\in A_0} (A_x\lolli B_{f(x)}).
  $
  Here, of course, $A_0\to B_0$ denotes the set of all functions from
  $A_0$ to $B_0$, and $A_x\lolli B_y$ denotes the exponential object
  in the monoidal closed category $\Mmx$. Note that in the definition
  of $C_f$, we have used the fact that $\Mmx$ has products.
\end{proposition}

We note that the category $\Setar$ of Section~\ref{sec-simple} is a
special case; indeed, if the initial monoidal category is $\Mm=\Set$,
then it is already closed, so we can take $\Mmx = \Set$, and we get
$\Mmxx\equivalent\Setar$.

\subsection{Properties of objects}\label{ssec-properties}

The concepts of parameter and state objects can be defined analogously
to Section~\ref{ssec-setar-object-properties}.  Namely, an object
$A\in \Mmxx$ is a {\em parameter object} if each fiber is the tensor
unit $I$, i.e., if $A=(X,(I)_{x\in X})$. An object $A\in \Mmxx$ is a
{\em state object} or {\em simple} if $A_0\iso 1$. Note that the full
subcategory of $\Mmxx$ of simple objects is equivalent to $\Mmx$.  An
object $A\in\Mmxx$ is an {\em M-object} if every fiber belongs to the
category $\Mm$ (regarded as a full subcategory of $\Mmx$). Thus,
M-objects denote families of objects of $\Mm$. Note that the full
subcategory of $\Mmxx$ of {\em simple M-objects} is equivalent to
$\Mm$. The terminology ``simple'' is justified by the fact that every
object $A=(X,(A_x)_x)$ of $\Mmxx$ can be written, essentially
uniquely, as a coproduct of simple objects.

More systematically, note that we have functors
$
\Set \catarrow{p} \Mmxx$ and $
\Mm \stackrel{i}{\hookrightarrow} \Mmx
\stackrel{j}{\hookrightarrow}\Mmxx$,
where $p(X)=(X,(I)_x)$, $i$ is the canonical inclusion, and
$j(A)=(1,(A))$. Then $A$ is a parameter object iff it is in the image
of the functor $p$, a simple object iff it is in the image of $j$, and
a simple M-object iff it is in the image of $j\circ i$.

\subsection{Some basic types and operations}

The coproducts of $\Mmxx$ permit us to construct a parameter object
$\Bool = I+I$. This object is equipped with morphisms
$\ttrue,\ffalse : I\to\Bool$ and an if-then-else construction, i.e.,
an operation mapping a pair of morphisms $f,g:A\to B$ to a morphism
$h:\Bool\x A\to B$ satisfying appropriate conditions.  Similarly,
there is an object $\Nat = (\N, (I)_{n})$, and indeed there is a
parameter object $p(X)$ corresponding to every set $X$, arising from
the functor $p:\Set\to\Mmxx$.

Assume the category $\Mm$ has some distinguished objects, say $\Bit$
and $\Qubit$, and some distinguished morphisms, for example
$H:\Qubit\to\Qubit$ and $\meas:\Qubit\to\Bit$. The latter are called
{\em gates}. Then there are corresponding objects and morphisms in
$\Mmxx$, arising from the embedding $\Mm\hookrightarrow\Mmxx$.

\subsection{Inductive types}

Recall that a functor $F$ on a category is {\em continuous} if it
preserves colimits of diagrams of the form $A_0\to A_1\to\ldots$. In a
category with colimits, every continuous functor has an initial
algebra. This is the basis for the categorical semantics of inductive
datatypes. Unfortunately, the category $\Mmxx$ only has coproducts,
and not necessarily all colimits, so we cannot in general expect
initial algebras of continuous functors to exist. Nevertheless, for
many functors of interest, the required initial algebras exist. For
example, consider the functor $F(X) = I + A\x X$. Its initial algebra
is the infinite coproduct $I+A+A\x A+A\x A\x A+\ldots$. We denote it
as $\List(A)$, the type of lists of $A$. An analogous construction
also works for other functors constructed from $+$ and $\x$.  Another
example is the object $\Nat$ of natural numbers, which arises as the
initial algebra of $F(X) = I+X$, and is also isomorphic to $\List(I)$.

\subsection{Boxing}\label{ssec-boxing}

\begin{proposition}
  The functor $p:\Set\to\Mmxx$, defined in
  Section~\ref{ssec-properties}, has a right adjoint
  $\flat : \Mmxx \to \Set$. It is given as follows, where
  $\Mmx(A,B)$ denotes a hom-set of the category $\Mmx$: 
  \[ \flat(X,(A_x)_{x\in X}) = \sum_{x\in X} \Mmx(I, A_x).
  \]
\end{proposition}

An important special case arises for simple M-objects $T$ and $U$.
In this case, we have
\begin{equation}\label{eqn-lolli}
  \flat(T\lolli U)\ \iso\ \Mmx(I, T\lolli U)\ \iso\ \Mmx(T, U)\
  \iso\ \Mm(T, U).
\end{equation}
In other words, $\flat(T\lolli U)$ is just a hom-set of the category
$\Mm$, i.e., a set of generalized circuits. We would like to be able
to use completed circuits as parameters in the construction of other
circuits, i.e., we would like there to be a parameter object whose
elements are circuits. Such an object is
$p(\Mm(T,U))\iso p(\flat(T\lolli U))$. This motivates the following
definition:

\begin{definition}\label{def-bang}
  The functor $\bang : \Mmxx \to\Mmxx$ is defined by
  $ \bang = p\circ \flat.
  $
  Since $p$ and $\flat$ are adjoints, the functor $\bang$ is a comonad
  on the category $\Mmxx$ {\cite{MacLane91}}. We call it the {\em
    boxing comonad}. It is equipped with a natural transformation
  $\force : \bang A \to A$ as well as a lifting operation
  \[
    \proofrule{f:\bang A\to B}
    {\lift(f) : \bang A\to \bang B.}
  \]
\end{definition}

\begin{remark}\label{rem-bang-hom}
  From {\eqref{eqn-lolli}}, we have an isomorphism
  $\bbox:\bang(T\lolli U)\to p(\Mm(T,U))$ for simple M-objects $T$ and
  $U$.  We denote its inverse by $\bunbox$.  It will also be
  convenient to consider an operation
  $\bapply : p(\Mm(T,U))\x T \to U$, which is definable from $\bunbox$
  using $\force$ and the monoidal closed structure.
\end{remark}

\begin{remark}\label{rem-parameter-lift}
  In the category $\Mmxx$, every parameter object $P$ is isomorphic to
  an object of the form $\bang A$. We can therefore generalize the
  lift operation to all parameter objects:
  \[
    \proofrule{f: P\to B}
    {\lift(f) : P\to \bang B.}
  \]
\end{remark}

\begin{theorem}
  The category $\Mmxx$, together with the adjunction given by
  $p:\Set\to\Mmxx$ and $\flat:\Mmxx\to\Set$, forms a linear-non-linear
  model in the sense of Benton {\cite{benton94mixed,Mellies09}}.
\end{theorem}

\subsection{Meta-operations on circuits}\label{ssec-meta}

In Remark~\ref{rem-meta}, we considered that the category $\Mm$ may be
equipped with additional meta-operations on generalized circuits, such
as
\[ \begin{array}{lll}
     \size &:& \Mm(T,U)\to \N,\\
     \print &:& \Mm(T,U) \to \Document,\\
     \invert &:& \Mm(T,T) \to \Mm(T,T)
   \end{array}
\]
Note that these operations are {\em external} to the category $\Mm$,
i.e., they are not morphisms of $\Mm$, but set-theoretic
functions. Using the boxing monad ``$\bang$'', these meta-operations
can be {\em internalized} in the category $\Mmxx$. Namely, given the
above operations, the following are morphisms of $\Mmxx$, where $T,U$
are simple M-objects:

\begin{center}
  $\begin{array}{lll}
  \bsize &=& \bang(T\lolli U)\catarrow{~\bbox~}
             p(\Mm(T,U)) \catarrow{~p(\size)~}
             p(\N) \catarrow{~\iso~}
             \Nat
  \\
  \bprint &=& \bang(T\lolli U)\catarrow{~\bbox~}
              p(\Mm(T,U)) \catarrow{~p(\print)~}
              p(\Document)
  \\
  \binvert &=& \bang(T\lolli T)\catarrow{~\bbox~}
               p(\Mm(T,T)) \catarrow{~p(\invert)~}
               p(\Mm(T,T)) \catarrow{~\bunbox~}
               \bang(T\lolli T)
\end{array}
$\end{center}

This precisely captures our intuition that ``boxed'' circuits are
concrete data that can be operated upon at circuit generation time.

\section{Towards a circuit description language}\label{sec-pl}

\subsection{Overview}

As explained in the introduction, we will design a programming
language for circuit description by making the language fit the
denotational model. We start with a brief overview of all the features
that will be added to the language. The formal definitions will follow
in Section~\ref{ssec-formal-qpl}.

Since the category $\Mmxx$ is symmetric monoidal closed with
coproducts, a standard linear lambda calculus with sum types can be
interpreted in it. Basic types such as $\Bool$, $\Bit$ and $\Qubit$
(the latter two if present in the category $\Mm$ of generalized
circuits) can also be added to the language, along with the associated
terms (such as $\ttrue$, $\ffalse$, an if-then-else construction, and
any basic gates that are present in the category $\Mm$). Moreover,
certain inductive types such as $\List(A)$ and $\Nat$
exist in the model and therefore can be added to the language.  The
language can further be equipped with a type operation ``$\bang$'' and
terms ``$\tlift$'', ``$\tforce$'', ``$\tbox{}$'', and ``$\tapply$'',
arising from their categorical counterparts in
Section~\ref{ssec-boxing}.

Certain types of the language will be designated as parameter types,
simple types, and/or M-types. Their interpretation in the model will
of course be parameter objects, simple objects, and/or M-objects,
respectively. In Quipper, M-types were called {\em quantum data
  types}, but that name does not seem appropriate in the context of
generalized circuits. We have not come up with a better name for them
and are stuck with ``M-types'' for now.

\begin{remark}\label{rem-descriptive}
  Our claim that the resulting programming language is a language for
  describing families of circuits is justified by the following
  observation. Suppose $\Phi\vdash N : T\lolli U$ is a valid typing
  judgement, where $\Phi$ is a parameter context (i.e., a context
  where each type is a parameter type), and $T$ and $U$ are simple
  M-types. Then the interpretation of this typing judgement will be a
  morphism $\sem{N} : p(X)\to \sem{T}\lolli\sem{U}$ of the category
  $\Mmxx$, where $p(X)=\sem{\Phi}$ is a parameter object and $\sem{T}$
  and $\sem{U}$ are simple M-objects. We have:
  \[ \Mmxx(p(X), \sem{T}\lolli\sem{U})
  ~\iso~
  \Set(X,\flat(\sem{T}\lolli\sem{U}))
  ~\iso~
  \Set(X,\Mm(\sem{T},\sem{U})),
  \]
  where the first isomorphism uses the fact that $\flat$ is the right
  adjoint of $p$, and the second isomorphism arises from
  {\eqref{eqn-lolli}}. Therefore, the interpretation of $N$ literally
  yields a function from $X$ to $\Mm(\sem{T},\sem{U})$, i.e., a
  parameterized family of generalized circuits. 
\end{remark}

\subsection{Labelled circuits}\label{ssec-labelled-circuits}

As explained in Section~\ref{ssec-category-Mm}, we assume that a
symmetric monoidal category $\Mm$ of {\em generalized circuits} has
been fixed once and for all. To make it more convenient for our
programming language to manipulate morphisms of $\Mm$, it is useful to
equip $\Mm$ with an additional {\em labelling structure}, which we now
define.

Let us assume a fixed given set $\Wiretypes$ of {\em wire types},
together with an interpretation function $\sem{-}:\Wiretypes\to|\Mm|$,
assigning an object of $\Mm$ to every wire type. In practical
examples, we will sometimes assume that the set $\Wiretypes$ contains
two wire types called $\Bit$ and $\Qubit$, but in general the set of
wire types is arbitrary. We often denote wire types by Greek letters
such as $\alpha$ and $\beta$.  Let $\Ll$ be a fixed countably infinite
set of {\em labels}, which we assume to be totally ordered. We often
denote labels by the letters $\ell$ or $\kay$. A {\em label context}
is a function from some finite set of labels to wire types.  We write
label contexts as $Q = \ell_1:\alpha_1,\ldots,\ell_n:\alpha_n$.  To
each label context $Q = \ell_1:\alpha_1,\ldots,\ell_n:\alpha_n$, we
associate the object $\sem{Q}=\sem{\alpha_1}\x\ldots\x\sem{\alpha_n}$
of $\Mm$, where $\ell_1<\ell_2<\ldots<\ell_n$. In case $Q=\emptyset$,
we set $\sem{Q} = I$.

\begin{definition}[Labelled circuits]\label{def-labelled}
  Let $\Mm$ be a given symmetric monoidal category, and let
  $\Wiretypes$, $\Ll$, and the function $\sem{-}:\Wiretypes\to|\Mm|$
  be given as above. The symmetric monoidal category $\Mm_{\Ll}$ of
  {\em labelled circuits} is defined as follows:
  \begin{itemize}
  \item The objects of $\Mm_{\Ll}$ are label contexts.
  \item A morphism $f:Q\to Q'$ in $\Mm_{\Ll}$ is by definition a
    morphism $f:\sem{Q}\to\sem{Q'}$.
  \end{itemize}
  Identities and composition are defined so that
  $\sem{-}:\Mm_{\Ll}\to\Mm$ is a full and faithful functor. We equip
  $\Mm_{\Ll}$ with the unique (up to natural isomorphism) symmetric
  monoidal structure making this functor symmetric monoidal.
\end{definition}

Note that if $Q$ and $Q'$ have disjoint domains, then
$Q\x Q'\iso Q\cup Q'$, i.e., the tensor of disjoint label contexts is
given by their union. Two label contexts can always be made disjoint
up to isomorphism by renaming their labels. We can visualize the
morphisms of $\Mm_{\Ll}$ as generalized circuits with labelled and
typed inputs and outputs, for example
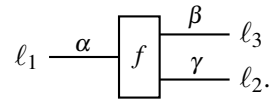
\[
\begin{qcircuit}[scale=0.6]
  \gridx{0}{2}{0.5};
  \gridx{2}{4}{0,1};
  \biggate{$f$}{2,0}{2,1};
  \leftlabel{$\ell_1$}{0,0.5};
  \rightlabel{$\ell_3$}{4,1};
  \rightlabel{$\ell_2$.}{4,0};
  \wirelabel{$\alpha$}{0.75,0.5};
  \wirelabel{$\beta$}{3.25,1};
  \wirelabel{$\gamma$}{3.25,0};
\end{qcircuit}
\]

\begin{remark}
  Although it was convenient to assume that the set of labels is
  totally ordered, Definition~\ref{def-labelled} is actually
  independent of the particular order chosen. Specifically, if we
  change the ordering of the labels, the resulting category $\Mm_{\Ll}$
  will be isomorphic to the original one. 
\end{remark}

\subsection{The syntax of Proto-Quipper-M}\label{ssec-formal-qpl}

\begin{table}
  \[
  \begin{array}{llll}
    \mbox{Types} 
    & A,B 
    & ::=
    & \alpha 
      \mid 0
      \mid A+B
      \mid I
      \mid A\x B
      \mid A\lolli B
      \mid \bang A
      \mid \nnat
      \mid \llist A
      \mid \Circ(T,U)
    \\
    \mbox{Parameter types} 
    & P,R 
    & ::=
    & 0
      \mid P+R
      \mid I
      \mid P\x R
      \mid \bang A
      \mid \nnat
      \mid \llist P
      \mid \Circ(T,U)
    \\
    \mbox{Simple M-types} 
    & T,U
    & ::=
    & \alpha
      \mid I
      \mid T\x U
  \end{array}
  \]
  \caption{The types of Proto-Quipper-M}
  \label{tab-types}
\end{table}

The {\em types} of Proto-Quipper-M are shown in Table~\ref{tab-types}.
Here, $\alpha$ ranges over the set $\Wiretypes$ of wire types.
$\Circ(T,U)$ is the type of generalized circuits with inputs $T$ and
outputs $U$. Denotationally, this type is the same as
$\bang(T\lolli U)$, but it will receive special treatment in the
operational semantics. 

\begin{table}
  \[
  \begin{array}{llll}
    \mbox{Terms}
    & M,N
    & ::=
    & x
      \mid \ell
      \mid c
      \mid \llet x = M\inn N
    \\ &&
    &
      \mid \Box_A M
      \mid \lleft_{A,B} M
      \mid \rright_{A,B} M
      \mid \caseof{M}{x}{N}{y}{P}
    \\ &&
    &
      \mid \unit
      \mid M;N
      \mid \pair{M,N}
      \mid \llet \pair{x,y} = M \inn N
      \mid \lam x^A.M
      \mid MN
    \\ &&
    &
      \mid \tlift M
      \mid \tforce M
      \mid \tbox{T} M
      \mid \tapply(M,N)
      \mid (\vell,C,\vell')
    \\
    \mbox{Label tuples}
    & \vell,\,\vkay
    & ::=
    & \ell
      \mid \unit
      \mid \pair{\vell,\vkay}
    \\
    \mbox{Values}
    & V,W
    & ::=
    & x
      \mid \ell
      \mid c
      \mid \lleft_{A,B} V
      \mid \rright_{A,B} V
      \mid \unit
      \mid \pair{V,W}
      \mid \lam x^A.M
      \mid \tlift M
      \mid (\vell,C,\vell')
  \end{array}
  \]
  \caption{The terms of Proto-Quipper-M}
  \label{tab-terms}
\end{table}

The {\em terms} of Proto-Quipper-M are shown in Table~\ref{tab-terms}.
Here, $x$ ranges over a countable set of {\em variables}, $\ell$
ranges over the set $\Ll$ of labels, and $c$ ranges over a given set
of {\em constants}. We assume that these sets are pairwise disjoint. A
label is a symbolic representation of a circuit state: it is a pointer
to an output of a labelled circuit. Constants have a fixed
interpretation, which depends on the chosen category $\Mm$. For
example, it will be convenient to assume constants for built-in gates
such as $H:\Qubit\lolli\Qubit$. If the category $\Mm$ is equipped with
meta-operations as in Remark~\ref{rem-meta}, then these can be
represented by constants such as $\tsize : \Circ(T, U)\to \nnat$,
$\tinvert : \Circ(T,U)\to\Circ(U,T)$, and so on.  Most of the other
terms follow standard lambda calculus conventions. The operator
$\Box_A:0\to A$ represents the unique function from the empty type
to any type $A$. The term $(\vell,C,\vell')$ represents a {\em boxed
  circuit}, i.e., a value of type $\Circ(T,U)$. Specifically, $C$ is a
morphism of the category $\Mm_{\Ll}$, representing a generalized
circuit, and $\vell$ and $\vell'$ are label tuples, establishing an
interface between the inputs and outputs of $C$ and the types $T$ and
$U$. Terms of the form $(\vell,C,\vell')$ are not intended to be
written directly by users of the programming language. Rather, such
terms represent values that are computed by the programming language.

\begin{table}
  \[
  \infer[\rulename{var}]{\Phi, x:A; \emptyset \vdash x:A}{}
  \qquad
  \infer[\rulename{label}]{\Phi; \ell:\alpha \vdash \ell:\alpha}{}
  \qquad
  \infer[\rulename{const}]{\Phi; \emptyset \vdash c:A_c}{}
  \]\[
  \infer[\rulename{abs}]{\Gamma; Q\vdash \lam x^A.M : A\lolli B}{
    \Gamma, x:A; Q\vdash M:B
  }
  \qquad
  \infer[\rulename{app}]{\Phi, \Gamma_1, \Gamma_2; Q_1, Q_2 \vdash MN : B}{
    \Phi,\Gamma_1; Q_1 \vdash M : A\lolli B
    \quad
    \Phi,\Gamma_2; Q_2 \vdash N : A
  }
  \]\[
  \infer[\rulename{lift}]{\Phi; \emptyset \vdash \tlift M : \bang A}{
    \Phi; \emptyset \vdash M : A
  }
  \qquad
  \infer[\rulename{force}]{\Gamma; Q\vdash \tforce M : A}{
    \Gamma; Q\vdash M : \bang A
  }
  \]\[
  \infer[\rulename{box}]{\Gamma; Q\vdash \tbox{T} M : \Circ(T,U)}{
    \Gamma; Q\vdash M : \bang(T\lolli U)
  }
  \qquad
  \infer[\rulename{apply}]{\Phi, \Gamma_1, \Gamma_2; Q_1, Q_2 \vdash \tapply(M,N) : U}{
    \Phi,\Gamma_1; Q_1 \vdash M : \Circ(T,U)
    \quad
    \Phi,\Gamma_2; Q_2 \vdash N : T
  }
  \]\[
  \infer[\rulename{circ}]{\Phi; \emptyset\vdash (\vell,C,\vell') : \Circ(T,U)}{
    \emptyset;Q\vdash \vell : T
    \quad
    \emptyset;Q'\vdash \vell' : U
    \quad
    C\in\Mm_{\Ll}(Q, Q')
  }
  \]
  \caption{The typing rules of Proto-Quipper-M (excerpt)}
  \label{tab-rules}
\end{table}

A {\em variable context} is a function from a finite set of variables
to types. We write a variable context as
$\Gamma = x_1:A_1,\ldots, x_n:A_n$.  A variable context in which all
types are parameter types is called a {\em parameter context}; we
sometimes denote parameter contexts by $\Phi$. A {\em label context}
was already defined in Section~\ref{ssec-labelled-circuits}, and is a
function from a finite set of labels to wire types. As usual, we write
$\Gamma,\Delta$ to denote the union of contexts, provided that they
have disjoint domain.

A {\em typing judgement} is of the form $\Gamma;Q\vdash M:A$, and it
informally means that if the variables and labels have the types
declared in $\Gamma$ and $Q$, then $M$ is well-typed of type $A$. A
selection of the typing rules for Proto-Quipper-M is shown in
Table~\ref{tab-rules}.  Most of the typing rules resemble the standard
rules for a linear lambda calculus; we will comment on a few
particular features of the type system. Note that in the typing rules,
$\Phi$ stands for a parameter context, whereas $\Gamma$ denotes an
arbitrary variable context (which can contain both parameter types and
non-parameter types). There is no formal distinction between these two
kinds of contexts, so it is entirely possible to have a type
derivation where a given type is part of $\Phi$ in one rule and part
of $\Gamma$ in another.  Let us call a variable whose type is not a
parameter type a {\em linear} variable.  The type system enforces that
labels and linear variables are used exactly once, whereas parameters
may be used any number of times or not at all.  In the typing rule for
constants, we have assumed that each constant $c$ is equipped with a
fixed type $A_c$.

\subsection{Categorical semantics}

Because of the preparatory work we did in Section~\ref{sec-mmxx}, the
categorical semantics of Proto-Quipper-M is now straightforward. The
semantics associates to each type $A$ an object $\sem{A}$ of the
category $\Mmxx$ in the obvious way: the interpretation $\sem{\alpha}$
of a wire type is assumed given, and each connective is interpreted as
``itself'', for example, $\sem{0}=0$, $\sem{A+B}=\sem{A}+\sem{B}$, and
so on. We also set $\sem{\Circ(T,U)}=p(\Mm(\sem{T},\sem{U}))$.  If
$\Gamma=x_1:A_1,\ldots,x_n:A_n$ is a typing context, we write
$\sem{\Gamma}=\sem{A_1}\x\ldots\x\sem{A_n}$.

Next, we associate a morphism
$\sem{\Gamma;Q\vdash M:A} : \sem{\Gamma}\x\sem{Q}\to\sem{A}$ to each
valid typing judgement. By abuse of notation, we sometimes denote this
simply as $\sem{M}$. We assume that each constant $c$ of type $A_c$ is
interpreted by a given fixed morphism $\sem{c}:I\to \sem{A_c}$. The
interpretation of type derivations is defined by induction on the
typing rules. For space reasons, we only show part of the
interpretation in Table~\ref{tab-sem-terms}. The interpretation uses
maps $\terminal:P\to I$ and $\diagonal:P\to P\x P$, which exist in the
category $\Mmxx$ whenever $P$ is a parameter object. Each judgement of
the form $Q\vdash\vell:T$ induces an isomorphism
$\sem{\vell}:\sem{Q}\to\sem{T}$, from which we can define a morphism
$f(\vell,C,\vell') = \sem{\vell'}\circ C\circ\sem{\vell}^{-1} :
\sem{T}\to\sem{U}$. This is used in the last rule in
Table~\ref{tab-sem-terms}.

\begin{table}
  \[ 
  \begin{array}{rcl}
    \sem{\Phi, x:A; \emptyset \vdash x:A} 
    &=& \sem{\Phi}\x\sem{A}\catarrow{\terminal\x\id}I\x\sem{A}\catarrow{\iso}A 
    \\
    \sem{\Phi; \ell:\alpha \vdash \ell:\alpha} 
    &=& \sem{\Phi}\x\sem{\alpha}\catarrow{\terminal\x\id}I\x\sem{\alpha}\catarrow{\iso}\alpha 
    \\
    \sem{\Phi; \emptyset \vdash \tlift M : \bang A}
    &=& \sem{\Phi}\catarrow{\lift\sem{M}}\bang\sem{A}
    \\
    \sem{\Gamma; Q\vdash \tbox{T} M : \Circ(T,U)}
    &=& \sem{\Gamma}\x\sem{Q}\catarrow{\sem{M}}\bang(\sem{T}\lolli\sem{U})\catarrow{\iso}p(\Mm(\sem{T},\sem{U}))
    \\
    \sem{\Phi, \Gamma_1, \Gamma_2; Q_1, Q_2 \vdash \tapply(M,N) : U}
    &=& \sem{\Phi}\x\sem{\Gamma_1}\x\sem{\Gamma_2}\x\sem{Q_1}\x\sem{Q_2}
        \\ &&\quad
        \catarrow{\diagonal\x\id}
        \sem{\Phi}\x\sem{\Phi}\x\sem{\Gamma_1}\x\sem{\Gamma_2}\x\sem{Q_1}\x\sem{Q_2}
        \\ &&\quad
        \catarrow{\iso}
        (\sem{\Phi}\x\sem{\Gamma_1}\x\sem{Q_1})\x
        (\sem{\Phi}\x\sem{\Gamma_2}\x\sem{Q_2})
        \\ &&\quad
        \catarrow{\sem{M}\x\sem{N}}
        p(\Mm(\sem{T},\sem{U}))\x\sem{T}
        \catarrow{\bapply}
        \sem{U}
    \\
    \sem{\Phi; \emptyset\vdash (\vell,C,\vell') : \Circ(T,U)}
    &=& \sem{\Phi} \catarrow{\terminal} I \catarrow{p(f(\vell,C,\vell'))}
        p(\Mm(\sem{T},\sem{U}))
  \end{array}
  \]
  \caption{The interpretation of type derivations (excerpt)}
  \label{tab-sem-terms}
\end{table}

\section{Operational semantics}\label{sec-operational}

\subsection{Evaluation rules}

We define the operational semantics of Proto-Quipper-M as a big-step
semantics {\cite{Kahn87}}. A {\em configuration} is a pair $(C,M)$ of
a labelled circuit $C$ and a term $M$. Recall that a labelled circuit
is, by definition, a morphism of the category
$\Mm_{\Ll}$. Intuitively, $C$ is the circuit ``currently being
constructed'' when the term $M$ is run.  A configuration is a {\em
  value configuration} if $M$ is a value. Evaluation takes the form of
an {\em evaluation relation} $(C,M)\eval(C',V)$. Its intuitive meaning
is: when the term $M$ is evaluated in the context of a partially
constructed circuit $C$, then it produces a circuit $C'$ (obtained
from $C$ by appending zero or more gates) and a value $V$. We also
define an {\em error relation} $(C,M)\eval\Error$, meaning that the
evaluation of $M$ in the context of the circuit $C$ produces a
run-time error. Examples of run-time errors are:
\begin{itemize}
\item {\em Run-time type errors.} For example, evaluating an
  application $MN$, where $M$ is not a function, or a projection
  $\pi_1 M$, when $M$ is not a pair.
\item {\em Unbound variable or label.} For example, using a variable
  $x$ that has not been defined, or trying to append a gate to a wire
  $\ell$ that does not exist.
\item {\em Cloning errors.}  For example, trying to append a $2$-input
  gate to wires $\ell$ and $\ell'$, where $\ell=\ell'$.
\end{itemize}

\begin{table}
  \[ 
  \infer{(C,x)              \eval \Error}{}
  \qquad
  \infer{(C,\ell)           \eval (C,\ell)}{}
  \qquad
  \infer{(C,c)              \eval (C,c)}{}
  \]\[
  \infer{(C,\lam x.M)                \eval (C,\lam x.M)}{}
  \]\[
  \infer{(C,MN)                      \eval (C''',W)}{
    (C,M)\eval (C',\lam x.M')
    \quad
    (C',N)\eval (C'',V)
    \quad
    (C'',M'[V/x])\eval (C''',W)
  }
  \qquad
  \infer{(C,MN)                      \eval \Error}{
    (C,M)\eval \otherwise
  }
  \]\[
  \infer{(C,\tlift M)                \eval (C,\tlift M)}{}
  \qquad
  \infer{(C,\tforce M)               \eval (C'', V)}{
    (C,M)\eval (C',\tlift M')
    \quad
    (C',M')\eval (C'',V)
  }
  \qquad
  \infer{(C,\tforce M)               \eval \Error}{
    (C,M)\eval \otherwise
  }
  \]\[
  \infer{(C,\tbox{T} M)               \eval (C', (\vell,D,\vell'))}{
    (C,M)\eval(C',\tlift N)
    \quad
    \freshlabels(T) = (Q,\vell)
    \quad
    (\id_Q, N\vell)\eval (D, \vell')
  }
  \]\[
  \infer{(C,\tbox{T} M)               \eval \Error}{
    (C,M)\eval(C',\tlift N)
    \quad
   \freshlabels(T) = (Q,\vell)
    \quad
    (\id_Q, N\vell)\eval \otherwise
  }
  \qquad
  \infer{(C,\tbox{T} M)               \eval \Error}{
    (C,M)\eval\otherwise
  }
  \]\[
  \infer{(C,\tapply(M,N))            \eval (C''', \vkay')}{
    (C,M)\eval (C',(\vell,D,\vell'))
    \quad
    (C',N)\eval(C'',\vkay)
    \quad
    \append(C'',\vkay,\vell,D,\vell') = (C''',\vkay')
  }
  \]\[
  \infer{(C,(\vell,D,\vell'))        \eval (C,(\vell,D,\vell'))}{}
  \]
\caption{The big-step semantics (excerpt)}
\label{tab-eval}
\end{table}

The evaluation and error relations are defined recursively. A
selection of the evaluation rules are shown in Table~\ref{tab-eval},
using some notations which we now explain. As usual, $M[V/x]$ denotes
capture-avoiding substitution, i.e., the result of replacing the
variable $x$ by the value $V$ in the term $M$.  In the hypotheses of
several rules, we have used the notation ``$(C,M)\eval\otherwise$''.
This is an abbreviation for $(C,M)\eval(C',W)$, where $W$ is not of
one of the explicit forms mentioned in a previous rule for the same
configuration. For example, in the rules for $\tforce$,
$(C,M)\eval\otherwise$ means $(C,M)\eval(C',W)$ where $W$ is not of
the form $\tlift M'$. All such ``$\otherwise$'' cases yield run-time
type errors.

Most of the evaluation rules are those of a standard call-by-value
lambda calculus. The rules that do all of the interesting ``work'' of
Proto-Quipper-M are those for ``$\tbox{}$'' and ``$\tapply$''. Namely,
these rules are responsible for the construction of circuits.  They
rely on two auxiliary functions, which we now explain. The rules for
``$\tbox{}$'' use a function $\freshlabels$.  Given a simple M-type $T$,
the operation $\freshlabels(T)$ returns a pair $(Q,\vell)$ of a label
context and a label tuple such that $Q\vdash\vell:T$. Moreover, the
labels in $\vell$ are chosen to be {\em fresh}, which means that they
do not occur in $N$. 

The rule for ``$\tapply$'' uses a function $\append$, defined as
follows. First, let us say that two boxed circuits $(\vell,D,\vell)$
and $(\vkay,D',\vkay')$ are {\em equivalent}, in symbols
$(\vell,D,\vell)\iso(\vkay,D',\vkay')$, if they only differ by a
renaming of labels. Given labelled circuits $C:Q_0\to Q_1$ and
$D:Q_2\to Q_3$ and label tuples $\vkay$, $\vell$, and $\vell'$, the
operation $\append(C,\vkay,\vell,D,\vell')$ finds $D'$ and $\vkay'$
such that $(\vkay,D',\vkay')\iso(\vell,D,\vell')$, and such that the
labels in $\vkay'$ are fresh. It returns $(C',\vkay')$, where $C'$ is
the labelled circuit obtained by connecting the inputs of $D'$ to the
matching outputs of $C$ like this:
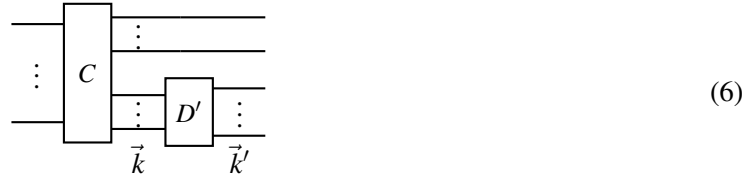
\begin{equation}\label{eqn-append}
\m{\begin{qcircuit}[scale=0.45]
  \gridx{0}{2}{0.2,3.1};
  \gridx{2}{5}{0,1,2.3,3.3};
  \gridx{5}{7.5}{-0.2,1.2,2.3,3.3};
  \biggate{$C$}{2,0}{2.5,3.3};
  \biggate{$D'$}{5,-0.1}{5.5,1.1};
  \clabel{$\vdots$}{0.75,1.80};
  \clabel{$\vdots$}{3.75,0.65};
  \clabel{$\vdots$}{6.75,0.65};
  \clabel{$\vdots$}{3.75,2.95};
  \wirelabel{\normalsize$\vkay$}{3.75,-1.5};
  \wirelabel{\normalsize$\vkay'$}{6.75,-1.5};
\end{qcircuit}
}
\end{equation}

\subsection{Safety properties}\label{ssec-safety}

The operational semantics satisfies the following safety properties: a
well-typed configuration never produces a run-time error, and if it
reduces to a value configuration, then the latter is well-typed of the
same type. To make this more precise, we first define a notion of
typing for configurations.

\begin{definition}
  Let $Q,Q'$ be label contexts, $(C,M)$ a configuration, and $A$ a
  type. We say that $(C,M)$ is {\em well-typed} with input labels $Q$,
  output labels $Q'$, and type $A$, in symbols
  $ Q\vdash (C,M):A;Q',
  $
  if there exists $Q''$ disjoint from $Q'$ such that
  $C:Q\to Q''\cup Q'$ and $\emptyset;Q''\vdash M:A$.
\end{definition}

\begin{proposition}[Subject reduction]
  If $Q\vdash (C,M):A;Q'$ and $(C,M)\eval(C',V)$, then
  $Q\vdash (C',V):A;Q'$.
\end{proposition}

\begin{proposition}[Error freeness]
  If $Q\vdash (C,M):A;Q'$, then $(C,M)\neval\Error$.
\end{proposition}

While the statement of these properties is succinct, the proofs are
intricate and require a number of auxiliary lemmas, which we omit.
Since our language does not have a recursion operator, we also have:

\begin{proposition}[Termination]\label{pro-termination}
  If $Q\vdash (C,M):A;Q'$, then there exists $(C',V)$ such that
  $(C,M)\eval(C',V)$.
\end{proposition}

\subsection{Soundness properties}

While the safety properties of Section~\ref{ssec-safety} relate 
the evaluation rules to the typing rules, the following {\em soundness
  properties} relate the evaluation rules to the categorical
semantics. We first extend the categorical semantics from well-typed
terms to well-typed configurations.

\begin{definition}
  To each well-typed configuration $Q\vdash (C,M):A;Q'$, we associate
  a morphism $\sem{(C,M)} : \sem{Q}\to\sem{A}\x\sem{Q'}$ of the
  category $\Mmxx$ as follows. By definition of well-typed
  configuration, there exists a unique $Q''$ such that $C:Q\to Q''\cup
  Q'$ and $\emptyset;Q''\vdash M:A$. Then
  \[ \sem{(C,M)} =
    \sem{Q}
    \catarrow{C}
    \sem{Q''\cup Q'}
    \catarrow{\iso}
    \sem{Q''}\x\sem{Q'}
    \catarrow{\sem{M}\x\id}
    \sem{A}\x\sem{Q'}
  \]
\end{definition}

\begin{proposition}[Soundness]
  If $Q\vdash (C,M):A;Q'$ is a well-typed configuration and
  $(C,M)\eval(C',V)$, then
  $\sem{(C,M)}=\sem{(C',V)} : \sem{Q}\to\sem{A}\x\sem{Q'}$.
\end{proposition}

The soundness property implies that the operational semantics
coincides with the categorical semantics. An important special case
arises if we have a closed term $M:\Circ(T,U)$. The categorical
meaning of $M$ is some generalized circuit
$\sem{M}:\sem{T}\to\sem{U}$. Operationally, the term $M$ evaluates to
some boxed circuit $(\vell,D,\vell')$, and soundness ensures that
$\sem{M}=\sem{(\vell,D,\vell')}$. Therefore, at {\em observable types}
such as $\Circ(T,U)$, our evaluation rules are a constructive
implementation of the categorical semantics. More generally, we have
the following adequacy property.

\begin{proposition}[Computational adequacy]
  If $\emptyset\vdash (C,M):A;\emptyset$ such that
  $\sem{(C,M)}=\sem{(C',V)}$ and $A$ is an observable type, then
  $(C,M)\eval(C',V)$ (possibly up to a renaming of labels in $V$).
\end{proposition}

As promised in Remark~\ref{rem-abstract}, the proofs of the soundness
and adequacy properties are independent of the choice of the category
$\Mmx$. This justifies not having made such a choice in the first
place.

\section{Conclusions and future work}

We systematically constructed a programming language for describing
families of generalized circuits by first giving a categorical model
and then defining the language to fit the model. The language has an
operational semantics, and we proved safety, soundness, and adequacy
properties showing that the computational and categorical meanings
coincide.

A software implementation of Proto-Quipper-M is in progress and almost
completed. In future work, we hope to extend the language to
encompass successively larger sets of features of the original
Quipper. For example, the current version of Proto-Quipper-M lacks a
general recursion scheme, so that all programs are terminating (see
Proposition~\ref{pro-termination}). Adding recursion to the
programming language is no problem at all, but how to add it to the
categorical model while preserving soundness and adequacy is an open
question.

Another question we hope to address in future work is the exact
relationship between Proto-Quipper-M and the version of Proto-Quipper
from Ross's thesis {\cite{RossThesis}}. While these two languages have
much in common, they differ in some important aspects: for example,
Ross's Proto-Quipper uses a subtyping relation $\bang A\subtype A$
instead of explicit ``$\tlift$'' and ``$\tforce$'' operators. This is
convenient for programmers, but giving a categorical semantics for
Ross's version of Proto-Quipper is left for future work.

\section{Acknowledgements}

This work was supported by the Natural Sciences and Engineering
Research Council of Canada (NSERC) and by the Air Force Office of
Scientific Research, Air Force Material Command, USAF under Award
No. FA9550-15-1-0331. This work was done in part while the authors
were visiting the Simons Institute for the Theory of Computing.

\bibliographystyle{eptcs}
\bibliography{pqmodel}

\begin{thebibliography}{10}
\providecommand{\bibitemdeclare}[2]{}
\providecommand{\surnamestart}{}
\providecommand{\surnameend}{}
\providecommand{\urlprefix}{Available at }
\providecommand{\url}[1]{\texttt{#1}}
\providecommand{\href}[2]{\texttt{#2}}
\providecommand{\urlalt}[2]{\href{#1}{#2}}
\providecommand{\doi}[1]{doi:\urlalt{http://dx.doi.org/#1}{#1}}
\providecommand{\bibinfo}[2]{#2}

\bibitemdeclare{inproceedings}{benton94mixed}
\bibitem{benton94mixed}
\bibinfo{author}{Nick \surnamestart Benton\surnameend} (\bibinfo{year}{1995}):
  \emph{\bibinfo{title}{A Mixed Linear and Non-Linear Logic: Proofs, Terms and
  Models (Extended Abstract)}}.
\newblock In: {\sl \bibinfo{booktitle}{Proceedings of the 8th Workshop on
  Computer Science Logic, CSL'94, Selected Papers}}, \bibinfo{series}{Springer
  Lecture Notes in Computer Science 933}, pp. \bibinfo{pages}{121--135},
  \doi{10.1007/BFb0022251}.

\bibitemdeclare{incollection}{Day70}
\bibitem{Day70}
\bibinfo{author}{Brian \surnamestart Day\surnameend} (\bibinfo{year}{1970}):
  \emph{\bibinfo{title}{On Closed Categories of Functors}}.
\newblock In: {\sl \bibinfo{booktitle}{Reports of the Midwest Category Seminar
  IV}}, {\sl \bibinfo{series}{Lecture Notes in Mathematics}}
  \bibinfo{volume}{137}, \bibinfo{publisher}{Springer}, pp.
  \bibinfo{pages}{1--38}, \doi{10.1007/BFb0060438}.

\bibitemdeclare{inproceedings}{GLRSV2013-pldi}
\bibitem{GLRSV2013-pldi}
\bibinfo{author}{Alexander \surnamestart Green\surnameend},
  \bibinfo{author}{Peter~LeFanu \surnamestart Lumsdaine\surnameend},
  \bibinfo{author}{Neil~J. \surnamestart Ross\surnameend},
  \bibinfo{author}{Peter \surnamestart Selinger\surnameend} \&
  \bibinfo{author}{Beno{\^\i}t \surnamestart Valiron\surnameend}
  (\bibinfo{year}{2013}): \emph{\bibinfo{title}{{Quipper}: a Scalable Quantum
  Programming Language}}.
\newblock In: {\sl \bibinfo{booktitle}{Proceedings of the 34th Annual ACM
  SIGPLAN Conference on Programming Language Design and Implementation, PLDI
  2013, Seattle}}, {\sl \bibinfo{series}{ACM SIGPLAN Notices}}
  \bibinfo{volume}{48(6)}, pp. \bibinfo{pages}{333--342},
  \doi{10.1145/2499370.2462177}.
\newblock \bibinfo{note}{Also available from \arxiv{1304.3390}}.

\bibitemdeclare{unpublished}{GLRSV2013-quipper}
\bibitem{GLRSV2013-quipper}
\bibinfo{author}{Alexander \surnamestart Green\surnameend},
  \bibinfo{author}{Peter~LeFanu \surnamestart Lumsdaine\surnameend},
  \bibinfo{author}{Neil~J. \surnamestart Ross\surnameend},
  \bibinfo{author}{Peter \surnamestart Selinger\surnameend} \&
  \bibinfo{author}{Beno{\^\i}t \surnamestart Valiron\surnameend}
  (\bibinfo{year}{2013}): \emph{\bibinfo{title}{The {Quipper} language}}.
\newblock \bibinfo{note}{Software implementation, available from
  \url{http://www.mathstat.dal.ca/~selinger/quipper/}}.

\bibitemdeclare{inproceedings}{Kahn87}
\bibitem{Kahn87}
\bibinfo{author}{Gilles \surnamestart Kahn\surnameend} (\bibinfo{year}{1987}):
  \emph{\bibinfo{title}{Natural Semantics}}.
\newblock In: {\sl \bibinfo{booktitle}{Proceedings of the 4th Annual Symposium
  on Theoretical Aspects of Computer Science (STACS 1987)}},
  \bibinfo{publisher}{Springer}, pp. \bibinfo{pages}{22--39},
  \doi{10.1007/BFb0039592}.

\bibitemdeclare{book}{MacLane91}
\bibitem{MacLane91}
\bibinfo{author}{S.~Mac \surnamestart Lane\surnameend} (\bibinfo{year}{1998}):
  \emph{\bibinfo{title}{Categories for the Working Mathematician}},
  \bibinfo{edition}{2nd} edition.
\newblock \bibinfo{series}{Graduate Texts in Mathematics},
  \bibinfo{publisher}{Springer}, \doi{10.1007/978-1-4757-4721-8}.

\bibitemdeclare{inproceedings}{Mellies09}
\bibitem{Mellies09}
\bibinfo{author}{Paul-Andr{\'e} \surnamestart Melli{\`e}s\surnameend}
  (\bibinfo{year}{2009}): \emph{\bibinfo{title}{Categorical semantics of linear
  logic}}.
\newblock In: {\sl \bibinfo{booktitle}{Interactive Models of Computation and
  Program Behaviour}}, {\sl \bibinfo{series}{Panoramas et
  Synth{\`e}ses}}~\bibinfo{volume}{27}, \bibinfo{publisher}{Soci{\'e}t{\'e}
  Math{\'e}matique de France}, pp. \bibinfo{pages}{1--196}.

\bibitemdeclare{inproceedings}{Paykin2017a}
\bibitem{Paykin2017a}
\bibinfo{author}{Jennifer \surnamestart Paykin\surnameend},
  \bibinfo{author}{Robert \surnamestart Rand\surnameend} \&
  \bibinfo{author}{Steve \surnamestart Zdancewic\surnameend}
  (\bibinfo{year}{2017}): \emph{\bibinfo{title}{QWIRE: A Core Language for
  Quantum Circuits}}.
\newblock In: {\sl \bibinfo{booktitle}{Proceedings of the 44th ACM SIGPLAN
  Symposium on Principles of Programming Languages}}, \bibinfo{series}{POPL
  2017}, \bibinfo{publisher}{ACM}, \bibinfo{address}{New York, NY, USA}, pp.
  \bibinfo{pages}{846--858}, \doi{10.1145/3009837.3009894}.

\bibitemdeclare{phdthesis}{RossThesis}
\bibitem{RossThesis}
\bibinfo{author}{Neil~J. \surnamestart Ross\surnameend} (\bibinfo{year}{2015}):
  \emph{\bibinfo{title}{Algebraic and Logical Methods in Quantum Computation}}.
\newblock Ph.D. thesis, \bibinfo{school}{Department of Mathematics and
  Statistics, Dalhousie University}.
\newblock \bibinfo{note}{Available from \arxiv{1510.02198}}.

\bibitemdeclare{incollection}{SV2009-qlambdabook}
\bibitem{SV2009-qlambdabook}
\bibinfo{author}{Peter \surnamestart Selinger\surnameend} \&
  \bibinfo{author}{Beno{\^\i}t \surnamestart Valiron\surnameend}
  (\bibinfo{year}{2009}): \emph{\bibinfo{title}{Quantum Lambda Calculus}}.
\newblock In \bibinfo{editor}{Simon \surnamestart Gay\surnameend} \&
  \bibinfo{editor}{Ian \surnamestart Mackie\surnameend}, editors: {\sl
  \bibinfo{booktitle}{Semantic Techniques in Quantum Computation}},
  chapter~\bibinfo{chapter}{4}, \bibinfo{publisher}{Cambridge University
  Press}, pp. \bibinfo{pages}{135--172}, \doi{10.1017/CBO9781139193313.005}.

\end{thebibliography}

\end{document}